\documentclass[12pt, letterpaper,]{article}
\usepackage[utf8]{inputenc}

\usepackage[numbers]{natbib}
\usepackage{bytefield}
\usepackage{rotating}
\usepackage{booktabs}
\usepackage{color}
\usepackage{wasysym}
\usepackage{amssymb}
\usepackage{etoolbox}
\usepackage{alltt}
\usepackage{hyperref}
\usepackage{float}
\usepackage{caption}
\usepackage{subcaption}
\usepackage{multirow}
\usepackage{enumerate}
\usepackage{listings}

\DeclareCaptionLabelFormat{andtable}{#1~#2  \&  \tablename~\thetable}

\usepackage{placeins}
\usepackage{balance}
\usepackage{epsfig}

\usepackage{color}
\usepackage{calc}
\usepackage{amstext}
\usepackage{graphicx}
\graphicspath{{../}}
\usepackage{makecell}

\usepackage{algorithm,algpseudocode}
\usepackage{multicol}
\usepackage{url}

\usepackage[underline=true]{pgf-umlsd}
\usetikzlibrary{positioning,calc}
\usetikzlibrary{decorations.pathreplacing}
\usepackage{placeins}

\usepackage{xparse}%

\usepackage{blindtext}

\usepackage{morewrites}

\usepackage{wrapfig}
\usepackage{xkeyval}%

\title{Multi-service Threats: Attacking and Protecting Network Printers and VoIP Phones alike
\\ \small{(Preprint accepted by \textit{Elsevier Internet of Things Journal} \\
DOI: \url{https://doi.org/10.1016/j.iot.2022.100507})}
}

\author{Giampaolo Bella$^1$, Pietro Biondi$^1$, Stefano Bognanni$^1$
\\
$^1$Dipartimento di Matematica e Informatica,\\
Universit\`{a} degli Studi di Catania. Catania, Italy
\\
email: giamp@dmi.unict.it, \\ pietro.biondi@phd.unict.it,\\  stefano.bognanni97@gmail.com}
\date{}

\begin{document}

\begin{titlepage}
\maketitle
\end{titlepage}

\begin{abstract}
Printing over a network and calling over VoIP technology are routine at present. This article investigates to what extent these services can be attacked using freeware in the real world if they are not configured securely. In finding out that attacks of high impact, termed the Printjack and Phonejack families, could be mounted at least from insiders, the article also observes that secure configurations do not appear to be widely adopted.
Users with the necessary skills may put existing security measures in place with printers, but would need novel measures, which the article prototypes, with phones in order for a pair of peers to call each other securely and without trusting anyone else, including sysadmins.
\end{abstract}

\section{Introduction}\label{sec:introduction}
The Internet of Things (IoT) is reality today but is deemed to become increasingly pervasive over time, realising the Internet of Everything. A few years ago, a breakthrough saw an attacker exploit a Samsung smart fridge to harvest a number of Gmail credentials~\cite{smartfridge}. The world has perhaps become accustomed to similar scandals ever since, such as the one with more than a million cameras turned into zombies to mount a massive DDoS attack~\cite{camerasvice} and the one with the burglars breaking into a house through its vulnerable smart deadbolt~\cite{iotscandal19}. Still, the overall 700\% rise of IoT malware attacks during the COVID-19 pandemic is very concerning~\cite{iotscandal21}.

Therefore, the motivation for our research stems from the verified assumption that the IoT gets bigger and bigger while still being insecure. We are aware that this may apply, in particular, also to printing over the network as well as to calling over VoIP technology if printers and phones are not configured securely. In trying to assess the large-scale impact of this statement on people, we formulate the following two research questions.
\begin{description}
\item[RQ1] \textit{Can printing over the network and calling over VoIP technology that are not configured securely be successfully and profitably attacked using freeware in a real-world scenario?}
\end{description}
This is a relevant question because, while a few tools specifically targeted at violating printers exist (\S\ref{sec:related}), it is not fully clear to what extent essential network sniffing and man-in-the-middle \textit{freeware} could be used to successfully defy network printers and VoIP phones alike, if these are not configured securely. Another important remark is that an answer can be sought \textit{in a real-world scenario}, thus not a laboratory mock-up, only from the standpoint of insider threats because we can experiment over our institutional network; by contrast, we lack the necessary authorisations and indemnities to probe real-world networks from the outside. It remains clear that a threat succeeding against network printers and VoIP phones could potentially be combined with any other external attack against the network.

\begin{description}
\item[RQ2] \textit{Can a user of a network secure their printing over the network and calling over VoIP technology from anyone, including from those with privileged access to the same network?}
\end{description}
This is the defence question logically paired up with the previous attack question. It is stated from the standpoint of a generic network user who wants to secure the printout they send to the institutional network printers as well as the VoIP calls they make to people they trust; notably, the question prescribes those uses to be secured \textit{from anyone}, thus including sysadmins with the potential, malicious activities they could carry out. We anticipate that attaining the level of protection that the question advocates is subject to the skills of the given user and their ability to reconfigure printers and phones. More importantly, it may not always be reached due to administrative restrictions that contributing network servers may be configured with.

To address these two research questions, we find out that, even using freeware such as \textit{nmap}, \textit{Ettercap}, \textit{Wireshark} and \textit{Python} programming, an attacker could seriously compromise network printers and VoIP phones. 
More precisely, the present article contributes two attack families, respectively termed Printjack, against network printers and Phonejack, against VoIP phones, with three attacks each.
Attacks number 1 are neither demonstrated nor original and see, respectively, the turning of a vulnerable printer or phone into a zombie to command and control towards a DDoS.
Attacks number 2 are fully demonstrated and represent an innovative instantiation of denial of service, respectively, to network printers, which use up their load of sheets, and to VoIP phones, which ring indefinitely. 
Also attacks number 3 are fully demonstrated and determine, respectively, unprecedented data breaches from a customary use of network printers and known yet simplified data breaches from a traditional use of VoIP phones.
All attacks are framed within a clearly stated threat model, and their likelihood is also evaluated from the standpoint of outsiders, namely any entity trying to attack from the outside of the network that hosts the target resources. Adequate protection measures against the presented attacks are provided, covering both existing as well as entirely novel measures at enforcing cryptographic tunnels.

This article is an abridged version of two workshop papers, one on Printjack~\cite{overtrustprinter} and one on Phonejack~\cite{paperVoIPBiondi}. It adds relevant research questions and their answers, a clear definition of the assumed threat model, an assessment of the attack likelihood from the outside of the network and a revised, fully-justified taxonomy for possible attack protection measures with respect to the stated threat model. The innovative material on top of what is already published is approximately 40\%.

The presentation is structured as follows. The state of the art is depicted through an outline of the relevant related work (\S\ref{sec:related}). The assumed threat model is then introduced and justified (\S\ref{sec:threatmodel}).
The treatment continues by describing the Printjack family of attacks against network printers (\S\ref{sec:printers}) and the Phonejack family against VoIP phones (\S\ref{sec:voipPhones}).  
All attacks are demonstrated from the standpoint of insiders, while the likelihood that they may also be executed by outsiders successfully is assessed separately (\S\ref{sec:likelihood}).
After that, the prose discusses possible attack protection measures against the two attack families, evaluating them with respect to the stated threat model. 
(\S\ref{sec:mitigations}). 
Some conclusions terminate the article (\S\ref{sec:conclusions}). The basics
of the essential VoIP protocols, SIP and RTP, are provided in Appendix~\ref{app:sip} for completeness.

\section{Related Work}\label{sec:related}
The state of the art features a few relevant contributions to the areas underlying the present article. These are briefly outlined here but are only indirectly useful to address our research questions --- this demonstrates the originality of our findings, particularly of the Printjack and Phonejack attack families.

\subsection{Printers}
The most eminent piece of research in the areas of printer security and privacy is due to M\"uller et al. \cite{expnetprinters}. They conduct a full-breadth vulnerability assessment and penetration testing session over a range of twenty commercial printers, comparing and contrasting a number of attacks on each of them. Their work is the first to note that raw 9100 port printing may be risky.

It must be mentioned that the work by M\"uller et al. also led to the development of the Printer Exploitation Toolkit (PRET), which is available on GitHub~\cite{pret}. However, we report that their tool did not work against our main testbed printer, a Lexmark MS620.
PRET is the newest and best developed of a small bunch of tools~\cite{toolwiki}, which did not cover our purposes either.

In the same year when the research findings by M\"uller et al. appeared, 2017, they were sided with breaking news reporting large-scale printer hacking somewhat for fun~\cite{printervice}, and the news was reiterated in 2018. The technical foundations behind the news remain vague. Moreover, it is not obvious to what extent the research findings inspired the events outlined in the news and, vice versa, whether the news partly ignited the researchers' investigations.

A recent study shows that also 3D printers, like traditional ones, may suffer security issues. In fact, McCormack et al. assess the security of networked 3D printers and their network implementations~\cite{3dprintsecurity}. They identified 8 types of vulnerabilities related to multiple types of DoS, lack of encryption and authentication, susceptibility to being spoofed, crash input, and unpatched vulnerabilities.

\subsection{VoIP}
Research in the field of VoIP is always a topic of central interest over time. The seminal contribution by McGann and Sicker looked at security threats and tools for SIP-based VoIP technology in 2005 \cite{Mcgann2005AnAO}. Their main conclusion was that testing tools do not always provide the coverage declared by the developers and may be difficult to install and configure properly. We find that work highly motivational but, regretfully, not followed up by much research. 

The most notable work is of 2010 by Keromytis \cite{angelostat}. He drew VoIP security statistics showing that 58\% of VoIP attacks are on Denial of Service, while 20\% are on eavesdropping, and hijacking. This reconfirms the importance of VoIP security measures and, in this vein, our work shows what can be done using modern freeware both for attack and defence purposes.

Next to the SIP-based VoIP technology tackled in the present paper, also TLS-based solutions such as SIPS \cite{SIPS} and SRTP \cite{SRTP} must be mentioned, with the extra ``S'' in their acronyms referring to security. In particular, SRTP uses asymmetric encryption to aim at authentication, integrity and confidentiality. It is clear that moving on to these technologies requires upgrading both the server and the phones.

\section{Threat Model}\label{sec:threatmodel}
The threat model we assume sees an attacker who conducts malicious activities raising the following threats:
\begin{itemize}
    \item  \textbf{Cybersecurity threats.} These are well known and target the illegitimate use of systems, resources or services, including denial of service.
    \item \textbf{Privacy threats.} These are increasingly relevant and target the breach of people's personal data, including access, disclosure and destruction of such data.
\end{itemize}

In general, such threats may either arise from the inside or from the outside of the network although, at an extreme, insiders may also attempt external threats:

\begin{itemize}
    \item \textbf{Internal Threats.}
The internal attack surface enables malicious access to any segment of the internal network by anyone who is registered within the given organisation.
The insider's specific activities include probing the internal network to do enumeration and information gathering, deriving useful knowledge to then mount attacks such as privilege escalation or lateral movement, also based on social-engineering tactics.
For example, the attacker may: use a network scanner such as \textit{nmap} \cite{nmap} to gather information with the aim of understanding the internal network architecture, the number and types of connected devices (including network printers and VoIP phones) and their open ports; exploit known vulnerabilities affecting the found systems and services or develop zero-day attacks for them. 
	
\item\textbf{External Threats.}
The external attack surface enables malicious access to any IP and ports that the boundary router of the given organisation exposes to the Internet. The outsider's specific activities include probing ranges of public IPs  to locate potential victims, exploring vulnerabilities of the exposed services to then mount attacks such as login-form injection and brute forcing. Other activities leverage social-engineering to convince users inside the organisation to click on fake links that install malicious software that virtually transforms the outsider in an insider.
For example, the attacker may: use an Internet-level scanner such as \textit{Shodan} \cite{shodan} to pinpoint specific addressed of Internet exposed resources;
(like insiders do internally)
exploit known vulnerabilities affecting the found systems and services or 
develop zero-day attacks for them; craft malicious links and attempt to dupe internal users; exploit vulnerabilities outside the target organisation, such as DNS spoofing and cache poisoning.
\end{itemize}

Our threat model particularly insists on internal threats. It means that it makes sense to conduct our experiments from within an internal network that also the attacker may access.
The selected network is one of our departmental VLANs, hence all findings reported below stem from this reference scenario.

Our threat model also allows external threats but we cannot address these because we do not have the necessary authorisations and indemnities to conduct the relevant activities in the real world, namely to probe an institutional network from the outside. 
However, to mitigate this limitation, a large-scale assessment will highlight the number of relevant IPs and ports that are publicly visible as a raising factor for the likelihood that our attacks may also be conducted from the outside of a network.

\section{Network printers}\label{sec:printers}
Our attacks to network printers leverage the common habit of using raw 9100 port printing.
A reference use case sees an insider scan their local network for available 9100 ports.
We define the Printjack family of attacks against network printers as follows~\cite{overtrustprinter}:
Printjack 1 attack, zombies for traditional DDoS (\S\ref{sec:zombie}); Printjack 2 attack, paper DoS (\S\ref{sec:paper}); Printjack 3 attack, privacy infringement (\S\ref{sec:privacy}).

\subsection{\textbf{Printjack 1 attack: zombies for traditional DDoS}}\label{sec:zombie}
Starting with network printers, it can be noted that there exist a number of documented vulnerabilities on various printers, which can be found on the Common Vulnerabilities and Exposures (CVE) database by the MITRE \cite{homemitre}. These observations motivate a daunting research question: how significant is the risk that worldwide printers get exploited to mount a massive DDoS attack? We argue that risk to be high hence worthy of mitigation, and provide supporting evidence for this argument below.

We take a stab at answering the question posed above by addressing the risk that a DDoS attack sourced from printers would take place.
There are a number of CVEs about printer vulnerabilities, precisely 217 can be found by querying the CVE database with keyword ``printer'' \cite{cveprinter} and 120 by querying it with keyword ``printers'' \cite{cveprinters}.
In particular, we observe that a few dozens of these allow for the remote execution of arbitrary commands or code. 
For example, CVE-2014-3741 ``\emph{allows remote attackers to execute arbitrary commands via unspecified characters in the lpr command}'' \cite{unavulnprinter}.

We contend that these findings, in combination with the potential for zero-day attacks, increase the likelihood of attacks on this field.
Thus, this type of attack can be used to build a zombie network that is useful for both internal and external threats.

\subsection{\textbf{Printjack 2 attack: paper DoS}}\label{sec:paper}
M\"uller et al. exhibit a proof of concept on how to mount a DoS on printers \cite{expnetprinters}. It keeps the PostScript interpreter of the printer busy forever by means of an infinite loop (based on an empty instruction and an empty exit condition). The researchers confirmed this attack on all their twenty tested printers but the HP LaserJet M2727nf, which automatically rebooted after ten minutes.

We note that raw port 9100 printing can be exploited to potentially exhaust the printing facilities of an institution. It can be done by abusing via the  9100 port any printer that becomes known through its IP address. An attacker would send repeated print jobs till the victim printer runs out of paper from all its paper trays. Looping on all institutional printers would then complete the attack. We conjecture that, in practice, a legitimate print attempt in front of a printer that processed all available paper (by printing something on each sheet and making it useless) would lead the employees to reload some paper trays. As an extreme, the institution would run out of paper
should the reloads persist before the attack is found and removed.

The Printjack 2 attack is of socio-technical nature because it is rooted in people's most obvious reaction to an aborted print attempt of theirs. It would be worth conducting field studies to verify our conjecture that people would feed their printers more and more paper unless they get their printout. This is beyond our present aims; by contrast, we provide a proof of concept implementation of the technical part below.

The technical part of the Printjack 2 attack can be easily implemented in Python as shown in Code \ref{code:script}. By looking at it from the inside out, we see a loop that sends each line, stored in {\tt textlines}, of a bot ASCII file {\tt bot.txt}, stored in {\tt textfile}, to a printer for a thousand times. The bot file could contain anything that the attacker may want to print in order to process and spoil paper sheets. The printer is identified via its IP address, and a socket connects to its 9100 port. The outermost loop ranges on the target IP addresses, which are read from file {\tt IPs.txt}. 

\begin{lstlisting}[language=python, caption={Python script for Printjack 2}, numberstyle=\tiny,  stepnumber=1, label={code:script}]
#file containing IPs of target printers
with open("IPs.txt") as f: 
 lines = f.read().splitlines() 
for ip in lines:
 #ascii file to be printed
 textfile = open("bot.txt", "r") 
 textlines = textfile.readlines()
 for count in range(0,1000): #number of print jobs
  s = socket.socket()
  s.connect((ip, 9100))
  for line in textlines:
   s.send(line+"\n")
   s.close()
\end{lstlisting}

\begin{figure}[ht]
	\begin{center}
		\includegraphics[width=0.95\linewidth]{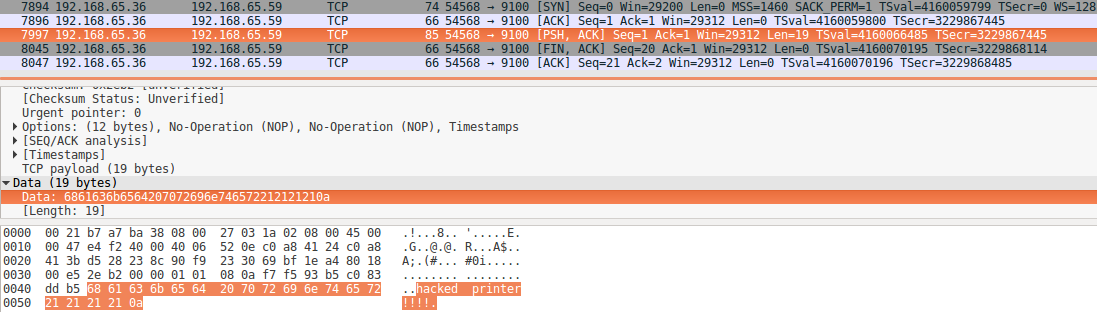}
	\end{center}
	\caption{The Printjack 2 attack monitored via Wireshark}\label{fig:sniffing1}
\end{figure}

We run our script on our institutional LAN. More in detail, we launched it from within the network, precisely from private IP address 192.168.65.36, towards a target printer of IP address 192.168.65.59. The printer exhausted its available paper by marking each sheet with the test phrase ``hacked printer!!!!''. Feeding it more paper would of course continue the paper abuse because the stated one thousand threshold had not been reached yet. We had to reset the printer manually to terminate the ignominy. Our experiment can be confirmed by observing the network traffic as sniffed by Wireshark \cite{wireshark}. The screenshot in Figure \ref{fig:sniffing1} highlights the appropriate TCP connection and the test phrase.

The Printjack 2 attack can be carried out on a specific target, or sprayed on different nodes connected to the network. Also, the attack can be carried out in an internal  threat scenario but we shall see that the likelihood that it may derive from external threats may not be negligible because worldwide available ports can be found, for example, through Shodan (\S\ref{sec:likelihood}).

\subsection{\textbf{Printjack 3 attack: privacy infringement}}\label{sec:privacy}

Internal threats can be represented as an attacker Mallory who sits on the same network as some target employee Alice. Whenever Alice sends a print job in the clear, Mallory could carry out a Man In The Middle (MITM) attack and eavesdrop the printed material, a clear infringement of Alice's privacy. Mallory could misbehave further, by publishing the intercepted material anonymously on the Internet, and produce a data breach.

A similar attack scenario sees a remote attacker Eve exploit one vulnerability into Alice's institutional network. It is state of the art to protect critical resources such as servers and databases by means of (strong) authentication. So, because Eve operates on the one node affected by the assumed vulnerability, those critical resources remain protected. By contrast, Eve could still perform the print job eavesdropping described above. Because printing is still common practice today, we cannot fully justify why data stored on a server would normally be protected and, by contrast, data sent off for printing would not.

The impact of such events would be very serious in our epoch, at least in the EU, where citizens' data protection is regulated by the General Data Protection Regulation (GDPR). With its 99 articles, the regulation empowers people with a number of rights to be exercised over their personal data as hosted by any data controller institution. The GDPR also stresses the responsibilities of the controller, for example article 5 paragraph 2 states that ``\emph{The controller shall be responsible for, and be able to demonstrate compliance with, paragraph 1 (`accountability').}'', with the mentioned paragraph 1 setting the requirement, among others, that data be ``\emph{processed in a manner that ensures appropriate security of the personal data, including protection against
	unauthorised or unlawful processing and against accidental loss, destruction or damage, using appropriate technical
	or organisational measures (`integrity and confidentiality').}''. Moreover, article 83 threatens ``\emph{administrative fines
	up to 20 000 000 EUR, or in the case of an undertaking, up to 4 preceding financial year, whichever is higher.}''.

Alice's institution has a great lot to worry about, equally because of Mallory's misconduct and because of Eve's.

Evidence seen in Figure \ref{fig:sniffing1} is valid also in the threat models embodied respectively by Mallory and Eve. In such cases, the visible traffic could be interpreted as Alice's, clearly intelligible, private data that Alice sent off for printing in a file intercepted by the attacker.

\begin{figure}[ht]
	\begin{center}
		\includegraphics[width=0.95\linewidth]{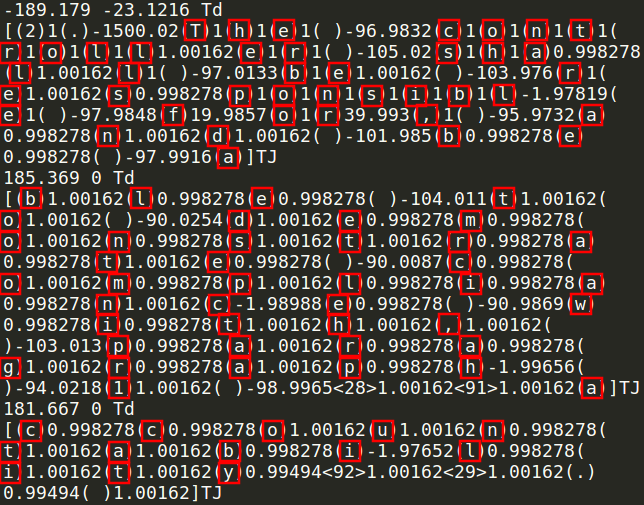}
	\end{center}
	\caption{Sniffing a PDF file (containing the GDPR) as printed from Linux}\label{fig:privacy2}
\end{figure}

We remark that raw port 9100 printing is massively used worldwide. For example, we observe that it is the default print method that the Common UNIX Printing System (CUPS) leverages, and that CUPS is vastly used in modern Linux distributions and Apple systems. As a demonstration, we used Ettercap \cite{ettercap} to interpose through sender and printer, then Wireshark to intercept the PDF file of the GDPR as from its official URL \cite{gdpr}. The outcome is intelligible with some decoding. The excerpt in Figure \ref{fig:privacy2} highlights in red the mentioned text of article 5 as intercepted over a print job sent from a Fedora 28 machine. It would be easy to implement a pretty-printing script.

Our print job sniffing experiments took a different course when the jobs were sent from an updated Windows 10 machine. While M\"uller et al. claim that Microsoft Windows printing architecture uses raw port 9100 printing by default \cite{expnetprinters}, our sniffing experiments yielded no comprehensible material. Although more experiments are needed to fully scrutinise this scenario, it would seem that 9100 no longer is the default printing port on Windows, thus supporting the claim that printing is more secure from Windows machines at present than from other systems.
Furthermore, regarding this scenario, we have tried several techniques~\cite{papercutWindows} including that of capturing the network traffic and replicating it to the printer in such a way as to have the ability to physically read the content from the printed sheet and not via the network.
Specifically, we sniffed a file sent to print via Windows 10 and noticed that the traffic was encrypted. So, we saved the stream and sent it back to the same printer with the aim of re-running the same print job, but despite ourselves, we failed in this experiment.

\begin{figure}[ht]
	\begin{center}
		\includegraphics[width=0.95\linewidth]{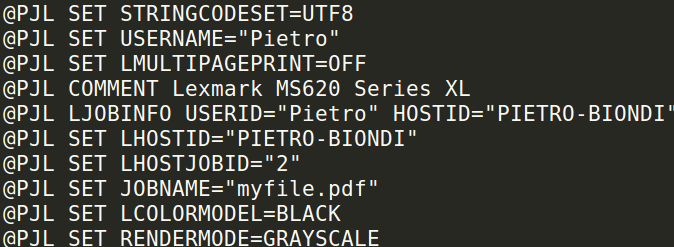}
	\end{center}
	\caption{Sniffing the metadata of a PDF file on Windows}\label{fig:privacy1}
\end{figure}

Nevertheless, we succeeded in intercepting the print job metadata on Windows. Figure
\ref{fig:privacy1} shows the metadata intercepted over port 65002, precisely fields {\tt USERNAME}, {\tt USERID}, {\tt HOSTID}, {\tt JOBNAME} as well as the printer model. Although this is less intrusive than accessing the contents of the printed file, it still counts as a data breach at least for the meaningful association of the file name to the user name. This claim rests on the socio-technical assumption that people give files meaningful names.

Also the Printjack 3 attack can be carried out towards a specific target, or sprayed on different nodes connected to the network.
The attack can only be carried out when the attacker has control of the network so as to be able to interpose, thus may only originate from internal threats.

\section{VoIP phones}\label{sec:voipPhones}
This section explains how insecure configurations of VoIP phones can be exploited through freeware, similarly to what was presented above about network printers. A reference use case sees an insider connect their laptop to the Ethernet cable disconnected from any VoIP phone. 
We define the Phonejack family of attacks against VoIP phones as follows~\cite{paperVoIPBiondi}: Phonejack 1 attack, zombies for DDoS (\S\ref{sec:voipzombie});
 Phonejack 2 attack, phone DoS (\S\ref{sec:voipdos});
Phonejack 3 attack, audio call eavesdropping (\S\ref{sec:privacyattack}).

\subsection{\textbf{Phonejack 1 attack: zombies for DDoS}} \label{sec:voipzombie}
As with printers, DoS continues to be one of the most dangerous attacks for VoIP.
Every model of a VoIP device may suffer zero-day or documented vulnerabilities. These could be exploited to carry out large-scale DDoS attacks against specific Internet targets. We conjecture but do not attempt such attack.
Therefore, we seek out to assess what vulnerabilities are known of VoIP systems. Querying the Common Vulnerabilities and Exposures (CVE) database of the MITRE \cite{homemitre} with keyword ``voip'' returns 110 CVE entries, some of which can be practically exploited on devices \cite{cvevoip}. Below we show two queries that can be searched, the result of which yields two CVEs, namely CVE-2014-3312, a Remote Command Execution (RCE)~\cite{CVE-2014-3312} and CVE-2014-3313, a Cross-Site Scripting (XSS)~\cite{CVE-2014-3313}. Both CVEs concern our testbed, therefore CISCO VoIP devices.

The impact of exploiting such vulnerabilities could be evaluated by the DDoS implications mentioned above. Querying Shodan \cite{shodan} may, in turn, help us assign a likelihood to such vulnerabilities, by informing us of how common the affected devices are. A ``cisco spa'' query returns 455 entries, which is a rather low outcome. A possible explanation is that VoIP devices are not left publicly visible, which is a commendable protection measure. However, a more general query, say ``asterisk'', returns 59.341 results, and would give a motivated attacker thousands of potential targets worth of further vulnerability assessment to seek VoIP exploitation. Of course, 2014 vulnerabilities have arguably been fixed ever since, but then we question whether updating phones falls into widespread security maintenance routine. We refrain from actively engaging into exploiting such vulnerabilities because this lies outside our research aims.

Finally, it must be recalled here that VoIP phones may also suffer undocumented, zero-day attacks.

\subsection{\textbf{Phonejack 2 attack: phone DoS}} \label{sec:voipdos}
While the exploitation of VoIP phones in a botnet might be somewhat ``traditional'', we also wonder to what extent phones themselves can become victims of a DoS activities. To assess such a vulnerability, we explore how to bombard a phone with tailored SIP packets, and observe that this can be successful.

By continuously sending packets to a target VoIP device, this can be exploited to ring indefinitely and crash eventually. We leverage Python multi-thread programming to overwhelm a test network of four VoIP devices and publish a video clip to demonstrate the audio experience of the attack~\cite{videophonejack}. To the best of our knowledge, Phonejack 2 is entirely innovative.

As preliminary operations, we configure the Asterisk server and the four Cisco VoIP phones so that they authenticate to the server. A phone ID, phone number, password and gateway must be manually entered on each phone. We assume that the phone number is public. An attacker may build a function to scan the local network and obtain the IPs and MAC addresses of the connected devices. Code \ref{code:netscanscript} shows a Python implementation of such a function. The \texttt{network} parameter represents the network to be scanned (e.g. 192.168.1.0/24).

\begin{lstlisting}[language=python, caption={Network scanning in Python}, numberstyle=\tiny,  stepnumber=1, label={code:netscanscript}]
def scanNetwork(network):
 hosts = []
 nm = nmap.PortScanner()
 out = nm.scan(hosts=network, arguments='-sP')
 for k, v in out['scan'].iteritems():
  if str(v['status']['state']) == 'up':
   hosts.append([str(v['addresses']['ipv4']),
   str(v['addresses']['mac'])])
return hosts 
\end{lstlisting}

We now make a call between two phones and record the network traffic through Wireshark \cite{wireshark}, as if run by an attacker. We apply a filter to extract the SIP packet that causes a ring, and save this packet in a file called \texttt{sipInvite.pcap}. This file contains information such as number and IP address of the recipient phone. We note that a phone does not check the SIP timestamp of a received packet but only that the recipient phone number of the packer corresponds to itself. Thus, a receiving phone only checks whether a packet is intended for itself. We also observe that flooding a phone with requests causes it to ring continuously, then crash and reboot. These observations guide our attack. We write a function (again in Python) to carve a SIP packet as we want. Our \texttt{flood\_DoS} function in Code \ref{code:floodDoS} takes an id, an IP address and a MAC address and calls the \texttt{tcprewrite} and \texttt{tcpreplay} commands. More specifically, \texttt{tcprewrite} takes the \texttt{sipInvite.pcap} file as input and modifies the fields containing the IP and MAC address of the packet. Finally, \texttt{tcprewrite} saves the new forged phone-ringing packet in a file called \texttt{newSipInvite.pcap}. After that, the \texttt{tcpreplay} command takes the \texttt{newSipInvite.pcap} file, sends it in loop to the phone and achieves the expected outcome: the phone rings for a few seconds, then crashes and reboots.

\begin{lstlisting}[language=python, caption={Phonejack 2 attack in Python}, numberstyle=\tiny,  stepnumber=1, label={code:floodDoS}]
def flood_DoS(id, IP, MAC):
 subprocess.call(['tcprewrite', '
 --dstipmap=192.168.1.18:'+IP,
 '--enet-dmac='+MAC,'--dlt=enet','--fixcsum',
 '--infile=sipInvite.pcap',
 '--outfile=newSipInvite'+id+'.pcap'])
 
 subprocess.Popen(['tcpreplay', '--intf1=eth0',
 '--loop=5',
 'newSipInvite'+id+'.pcap'])
return 
\end{lstlisting}

We then to carry out this attack in parallel on all our four devices. Code \ref{code:mainpy} shows a Python script that uses a thread for each phone. More precisely, the script scans the network using the \texttt{scanNetwork} function, builds a thread, gives it a job by means of the \texttt{flood\_DoS} function and starts it.

\begin{lstlisting}[language=python, caption={Parallelising the Phonejack 2 attack in Python}, numberstyle=\tiny,  stepnumber=1, label={code:mainpy}]
if __name__ == "__main__":
 hosts = scanNetwork(sys.argv[1])
 jobs = []
 for i in range(0, len(hosts)):
  IP=hosts[i][0]
  MAC=hosts[i][1]
  thr = threading.Thread(target=flood_DoS(i,IP,MAC))
  jobs.append(thr)
 for j in jobs:
  j.start()
 for j in jobs:
  j.join()
\end{lstlisting}

\begin{figure}[ht]
	\begin{center}
		\includegraphics[width=0.95\linewidth]{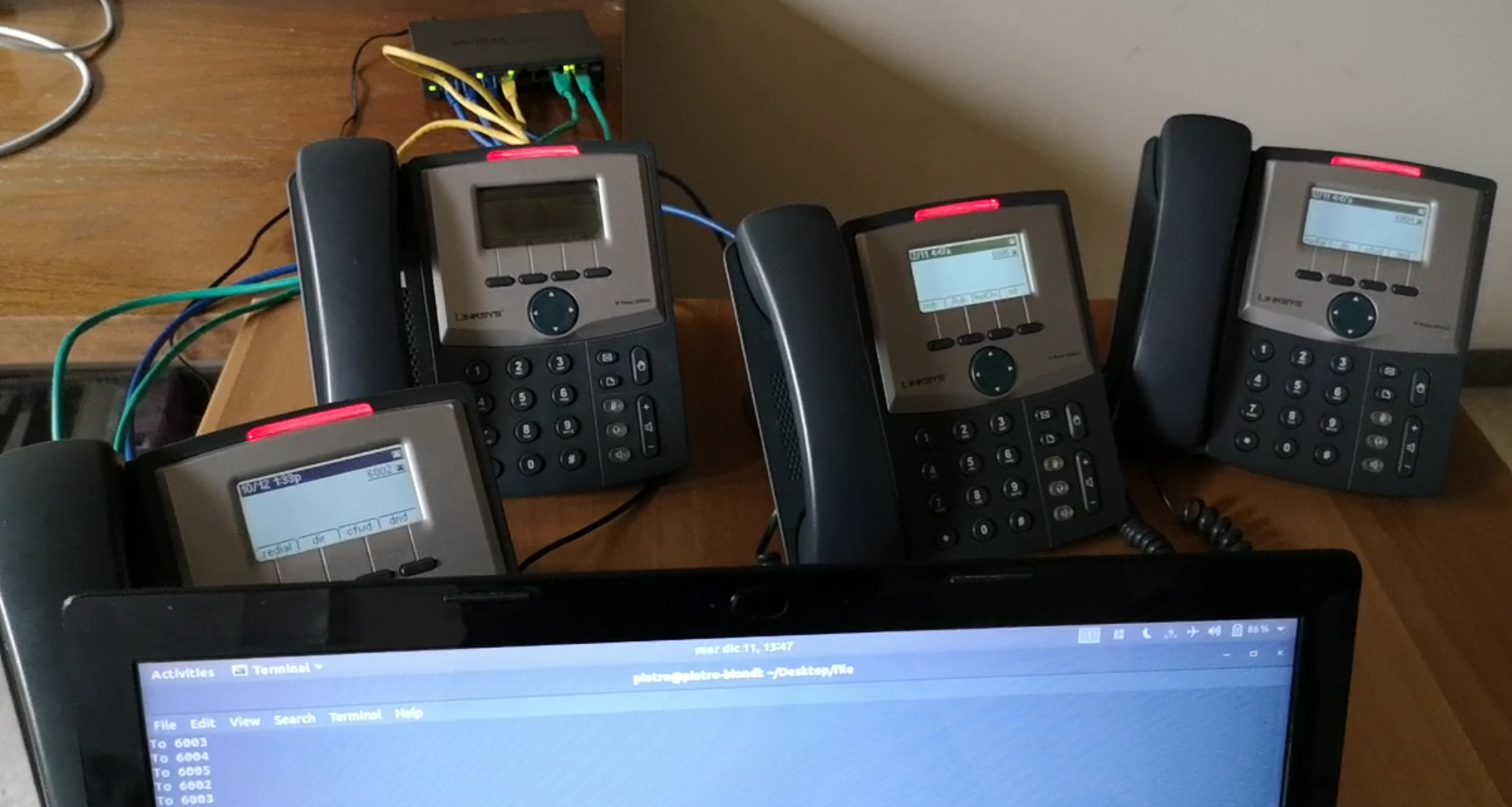}
	\end{center}
	\caption{Consequences of Phonejack 2}\label{fig:telphonejack2}
\end{figure}

Figure \ref{fig:telphonejack2} shows a laptop executing the Phonejack 2 attack against our four VoIP phones. Remarkably, all phones are ringing, as demonstrated by the red light on each of them. To better explain our results, we built a video clip \cite{videophonejack}.

It can be imagined that spraying this attack over the entire institution would have dramatic consequences. Not only would the calling service be hindered and ultimately zeroed, but the work environment would realistically become unbearable. 
Therefore, the Phonejack 2 attack can be carried out towards a specific target, or sprayed on different nodes connected to the internal network. Phonejack 2 is certain to work if mounted from the within the target network but seems unlikely to work from the outside because VoIP devices are often located behind a PBX server and are not reachable from the external network. However, unreachability from the outside is subject to proper configurations, a precondition that another assessment of ours (\S\ref{sec:likelihood}) raises concern about.

\subsection{\textbf{Phonejack 3 attack: audio call eavesdropping}} \label{sec:privacyattack}

Because packets are sent in the clear, we eavesdrop successfully and dump them to an audio file. This attack is based on the mentioned 2002 SANS observations \cite{sans} but it is intriguing that it solely relies on freeware.

Let us assume that Alice and Bob want to get in touch and that the attacker Eve is present in the same network. Since calls are made in the clear, Even could attempt sniffing a call between Alice and Bob, clearly infringing their privacy.

\begin{figure}[ht]
	\begin{center}
		\includegraphics[width=0.95\linewidth]{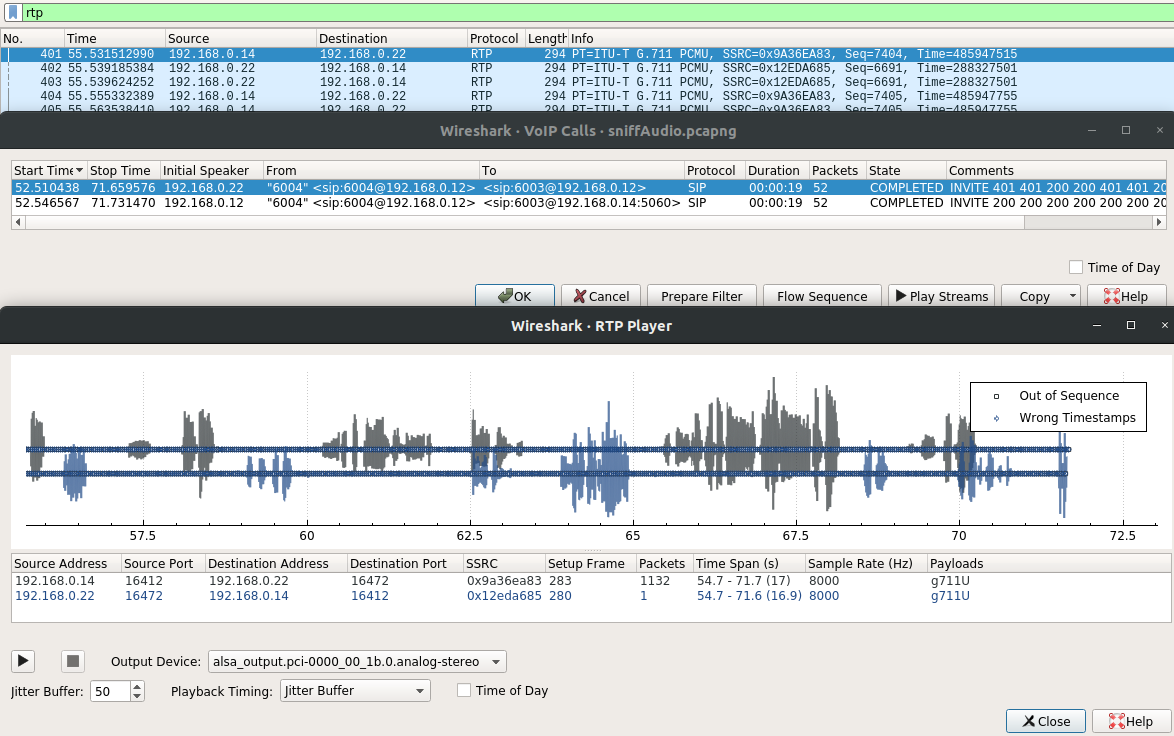}
	\end{center}
	\caption{Consequences of Phonejack 3}\label{fig:phonejack3}
\end{figure}

This conjecture can be demonstrated by taking the following steps. First, use Ettercap \cite{ettercap} to perform a Man in The Middle attack. Then, use a feature of Wireshark to listen to the audio flow of communication between two devices. Figure \ref{fig:phonejack3} shows the RTP traffic in the clear, as sniffed through Wireshark and played. To do this, we select an RTP packet, use the \texttt{Telephony} option and select the \texttt{VoIP Calls} feature. After that, we select one of the two streams and press \texttt{Play Stream}. Moreover, at the end of the call, we can export the audio track of the call as shown in our video clip \cite{videophonejack}.

Clearly, this attack could be leveraged to exfiltrate data also at the industrial espionage level. As noted above, it is not conceptually innovative but it is remarkable that we succeeded in carrying it out by using only freeware.

Finally, the Phonejack 3 attack can be carried out against specific targets connected to the internal network. Also Phonejack 3, like  Phonejack 2, can be carried out by an attacker from within the network to which the target VoIP phones are connected. It does not seem viable for external attackers due to the difficulties of interposing without hacking other network services such as DNS.

\section{Attack likelihood from the outside}\label{sec:likelihood}
A full assessment of the likelihood that the Printjack and Phonejack attacks may successfully be conducted from the outside of a target network is out of reach. Influencing factors are innumerable, mostly arising from general network vulnerabilities. These may be inherently technical, e.g. affect the network perimeter because a new exploit gets published against the institutional firewall/IDS, or socio-technical, e.g. affect the network users because social engineering strikes back.

However, we searched the number of ports traditionally associated with the printing and VoIP services that are also publicly visible through the Internet. Although we never probed those ports (as discussed above, \S\ref{sec:threatmodel}) hence cannot gauge to what extent they are configured securely, we recognise a significant number of such ports as a contributing factor for the likelihood that our attacks may succeed from the outside. To evaluate this factor, even a free student account for Shodan suffices, as we shall see below.

\subsection{Printjack attacks}
\begin{figure}
    \centering
    \includegraphics[width=0.95\linewidth]{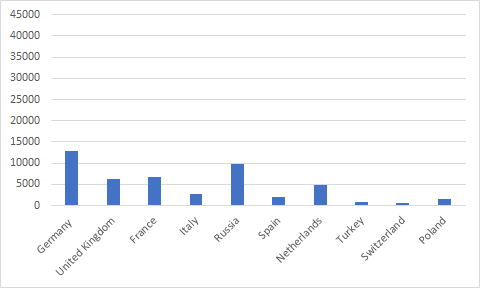}
    \caption{Graphical sorting of countries by responding 9100 port in February 2019}\label{fig:ist2019Printers}
\end{figure}

This Section discusses how common the bad practice of exposing public IP addresses over the Internet with a responding 9100 port is. We were surprised to find a few thousand occurrences in the authors' country, which arise through the Shodan query:
\begin{alltt}
	port:9100 country:"IT"
\end{alltt}

By varying the country identifier, Shodan continued to output unexpected results: Table \ref{tab:pil2019} sorts European countries by their 2019 Gross Domestic Profit (GDP) \cite{gdp} and reports the corresponding numbers of IPs with open 9100 ports, exposed to the Internet from that country. Data were gathered through Shodan in February 2019. It can be seen that, for example, the country with the highest GDP in 2019, Germany, also exposed the highest number of devices.
The histogram in Figure~\ref{fig:ist2019Printers} highlights that, however, the two rankings do not correlate any more. In particular, the country with the second highest number of visible ports, Russia, is number 5 in terms of GDP, and the country with the third highest number of visible ports, France, is number 3 in terms of GDP.

\begin{table}[ht]
\caption{IPs with responding 9100 port in February 2019 per country, sorted by country's GDP (2019)}\label{tab:pil2019} 
\begin{center}
	\begin{tabular}{|c|c|c|}
		\hline 
		GDP (2019) & Country & \makecell{IPs with responding\\ 9100 port in February 2019} \\ 
		\hline 
		1 & Germany & 12.891 \\ 
		\hline 
		2 & United Kingdom & 6.349 \\
		\hline 
		3 & France & 6.634 \\ 
		\hline 
		4 & Italy & 2.787 \\ 
		\hline 
		5 & Russia & 9.737 \\ 
		\hline
		6 & Spain & 2.088 \\ 
		\hline 
		7 & Netherlands & 4.934 \\
		\hline 
		8 & Turkey & 835 \\
		\hline 
		9 & Switzerland & 624 \\ 
		\hline
		10 & Poland & 1.425 \\ 
		\hline 
		 &  & Total = 48.304 \\ 
		\hline 
	\end{tabular}
\end{center}
\end{table}

In July 2021, we went back to Shodan and repeated the same experiments to verify whether the 2020 top-ten European countries by GDP were exposing a substantially varied number of openly reachable 9100 ports. The outcome is given in Table \ref{tab:pil2021}.
The list of countries by GDP is almost identical to the previous year's, with the only variation of Turkey and Switzerland swapping their places, so the list offers a good element of comparison through time.

It is remarkable that the total number of exposed 9100 ports increased by 150\%. 
All countries in fact experienced an increase, with Russia recording a remarkable record increase of 338\%. The country with the lowest increase was Italy, with only 7\%, while the average increase was of 106\%.
\begin{table}[ht]
\caption{IPs with responding 9100 port in July 2021 per country, sorted by country's GDP (2020)}\label{tab:pil2021} \begin{center}
	\begin{tabular}{|c|c|c|c|}
		\hline 
		GDP (2020) & Country & \makecell{IPs with responding\\ 9100 port in July 2021}  & Variation \\
		\hline 
		1 & Germany & 29.951 & +132\% \\ 
		\hline 
		2 & United Kingdom & 11.859 &  +87\% \\
		\hline 
		3 & France & 16.001 & +141\% \\ 
		\hline 
		4 & Italy & 2.982 & +7\% \\ 
		\hline 
		5 & Russia & 42.604 & +338\% \\ 
		\hline
		6 & Spain & 3.665 & +76\% \\ 
		\hline 
		7 & Netherlands & 8.273 & +68\% \\
		\hline 
		8 & Switzerland & 1.126 & +80\% \\
		\hline 
		9 & Turkey & 1.013 & +21\% \\ 
		\hline
		10 & Poland & 3.061 & +115\%   \\ 
		\hline 
		 &  & Total = 120.535 & +150\% \\
		\hline
	\end{tabular}
\end{center}
\end{table}
\begin{figure}
    \centering
    \includegraphics[width=0.95\linewidth]{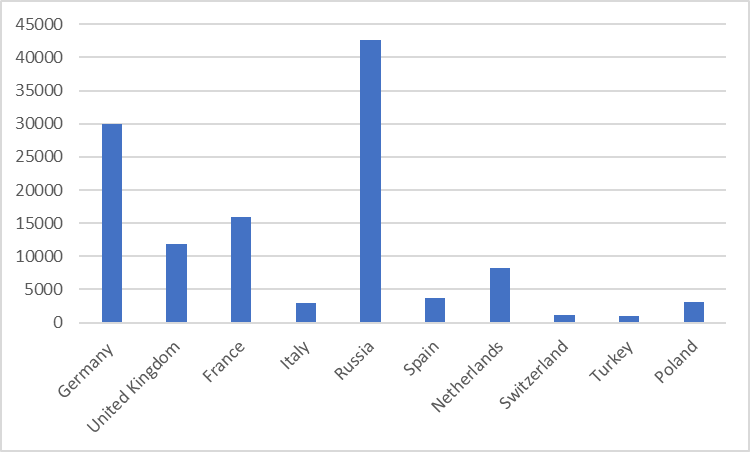}
    \caption{Graphical sorting of countries by responding 9100 port in July 2021}
    \label{fig:ist2021Printers}
\end{figure}
Figure~\ref{fig:ist2021Printers} offers a graphical highlight that the top three countries in terms of ports continue to be Russia, Germany and France, but the first two are now swapped, while their positions in the GDP sorting are unvaried, i.e., respectively 5, 3 and 1.

These findings are difficult to interpret fully. A correlation of potentially ill-configured printers that is worth of further investigation is with the COVID-19 pandemic. 
Also the correlation between the increasing use of remote cloud-printing services and raw 9100 printing is worth of a dedicated vulnerability assessment.

The remarkable increase of exposed 9100 ports may be interpreted a contributing factor to the likelihood that the Printjack attacks, Printjack 2 in particular, may succeed even by an outsider.
Of course, any service can be exposed via any port on one hand but, on the other, changing default port associations may not be common practice.

\subsection{Phonejack attacks}
\begin{figure}[ht]
    \centering
    \includegraphics[width=0.95\linewidth]{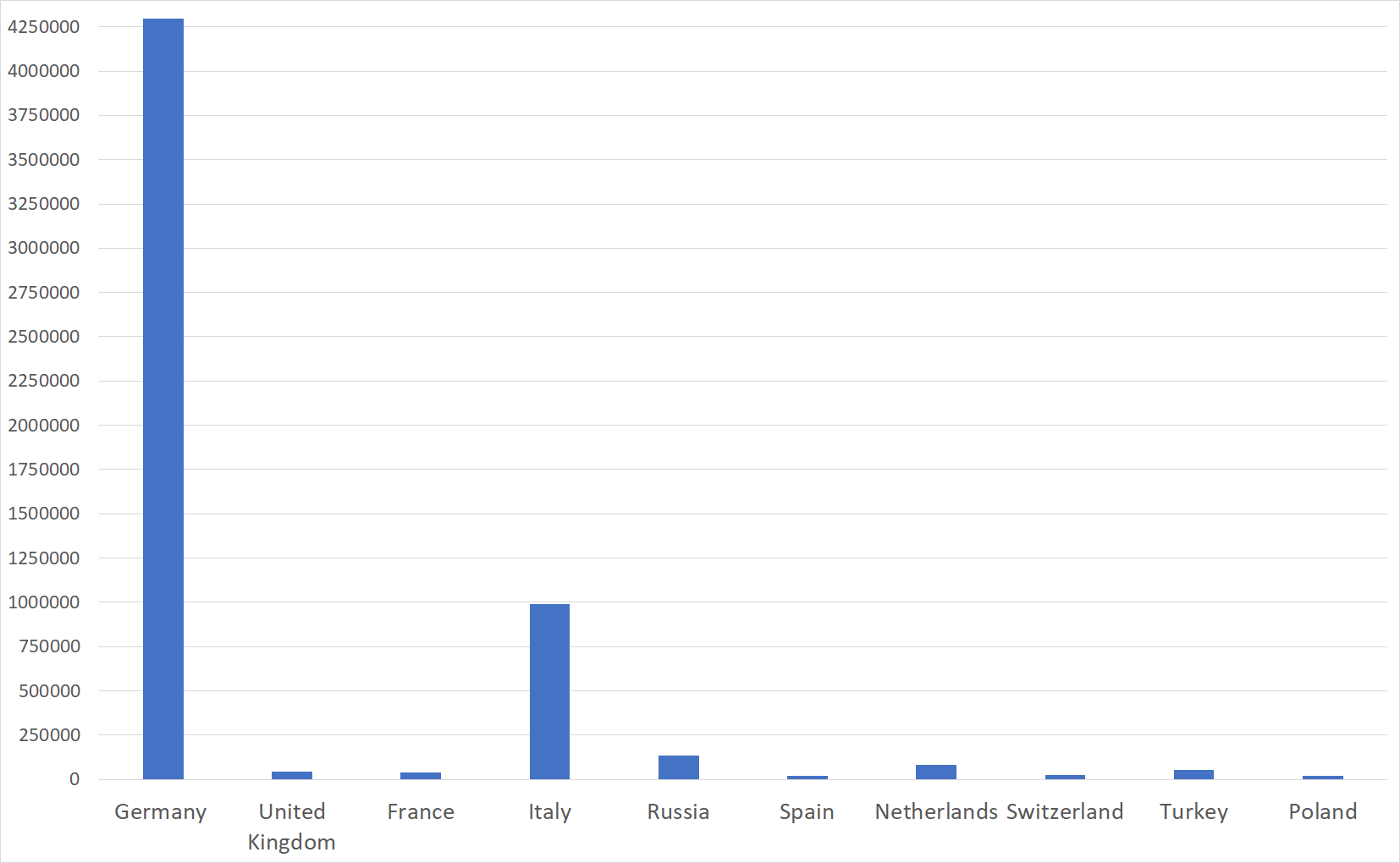}
     \caption{Graphical sorting of countries by responding 5060 port in July 2021}
    \label{fig:istoVoIP2021}
\end{figure}

Similar considerations apply to a likelihood assessment over the Phonejack attacks initiated from outside a target network. Therefore, also in this case we evaluated an objective factor, namely the international popularity of the visibility of the relevant network ports. Once more, the findings came through Shodan, and are reported in Table~\ref{tab:pilVoIP2021}, highlighting a total of over 5 million visible 5060 SIP ports. The histogram in Figure~\ref{fig:istoVoIP2021} shows that Germany is by far the country with the highest number of exposed ports, over 4 millions. Then, much later, comes Italy, with around 1 million, while other countries only expose a few tens of thousands each.

It is unfortunate that we do not have historical data to observe changes over time. However, we also searched for the number of visible 5061 ports, traditionally responding to SIP over TLS, and these turn out to be three orders of magnitude lower, only 3930 in total over the reference ten countries. 
These findings convince that the secure version of SIP is still much rarer than the traditional version, raising the likelihood that the Phonejack attacks, Phonejack 2 in particular, may succeed even by an outsider. 

It must be noted that, contrarily to the printer scenario, well-crafted SIP packets also require inclusion of the MAC address corresponding to the target IP as well as of the associated phone number. Shodan does not return that information, so the attacker will have to seek additional attack vectors, which may still be found by various means, including social-engineering. Moreover, the very payload that causes a phone to ring must be used, but it may be assumed to be public; flooding the phone with a different payload is likely to crash it and force it to reboot, a sort of ``silent, episodic DoS''.

\begin{table}[ht]
\caption{IPs with responding 5060 port in July 2021 per country, sorted by country's GDP (2020)}\label{tab:pilVoIP2021} 
\begin{center}
	\begin{tabular}{|c|c|c|}
		\hline 
		GDP (2020) & Country & \makecell{IPs with responding\\ 5060 port in July 2021} \\ 
		\hline 
		1 & Germany & 4.297.178 \\ 
		\hline 
		2 & United Kingdom & 42.012 \\
		\hline 
		3 & France & 38.699 \\ 
		\hline 
		4 & Italy & 992.085 \\ 
		\hline 
		5 & Russia & 133.163 \\ 
		\hline
		6 & Spain & 19.268 \\ 
		\hline 
		7 & Netherlands & 84.002 \\
		\hline 
		8 & Switzerland & 22.433 \\
		\hline 
		9 & Turkey & 53.535 \\ 
		\hline
		10 & Poland & 18.695 \\ 
		\hline 
		 &  & Total = 5.701.070 \\ 
		\hline 
	\end{tabular}
\end{center}
\end{table}

\section{Attack protection measures}\label{sec:mitigations}

Risk mitigation measures shaping up as protective measures are part of the risk assessment process for securing systems and devices in general, hence we also investigate how to counter the attacks reported above. 
The starting point comes from existing measures that could be put into place straight away. 

A few relevant security measures exist and, as we shall see, could be effective to some extent, but we contend that they are not widespread yet.
It is noteworthy, however, that the existing measures cannot always protect end-users, such as a company employee, from our threat model --- specifically, VoIP calls cannot be protected from malicious sysadmins, hence additional, novel measures must be prototyped.

\subsection{Existing measures}

\subsubsection{Network printers}
The best solution to thwart the Printjack attacks is to use IPSec in point-to-point mode, as more and more printers are supporting this protocol.

This application of IPSec ensures that the client sending the printout
gets authenticated to the printer, for example, via a pre-shared key that is manually entered in the client and also in the printer. 
The printer only accepts communications with authenticated clients and quits all other  attempts quickly, hence effectively countering resource exhaustion attacks such as Printjack 2.
Moreover, the client communicates with the IPSec responding printer through an encrypted tunnel, hence potential privacy breaches such as Printjack 3 are ruled out.

We expect that IPSec configurations are usually performed by sysadmins of host institutions, so a malicious administrator could realistically mount the Printjack attacks against end-users. 
However, every end-user, if sufficiently skilled, could in principle engage into redoing the IPSec configuration from scratch by entering secrets of own choice, then get protection from everyone.
%it is an operation that can be easily re-performed, through the printer manufacturer's guide, by any user in case they do not trust the system administrator.
%
To evaluate the likelihood that this happens, we tried various configurations of IPSec on both Microsoft and Linux systems. 
Our subjective evaluation is that IPSec configurations are very well documented for Microsoft systems thus reasonably easy to finalise for virtually everyone. By contrast, they are much more complicated to configure for Linux and may require sysadmin expertise.

It still remains to be seen how common it is for sysadmins to default to IPSec over institutional networks. We asked the question on various platforms, such as ``security.stackexchange'', ``telegram groups'' and ``quora''~\cite{quoraIPSec}, with awareness of the limits to the acceptability of the responses, and without claiming  statistically relevant findings. 
But most of the responses (even from sysadmins) pointed out that it is not very common to use IPSec to secure printers because unfortunately it is often assumed that the internal network is secure and that registered users will not harm it. This is a clear contradiction to widely recognised insider threats, as demonstrated in this article, raising preoccupation for the Printjack attacks.

Incidentally, a number of responses confirmed that printers are often misconfigured and become externally visible, reaffirming our assessment on the attack likelihood from the outside (\S\ref{sec:likelihood}).

\subsubsection{VoIP phones}
Security measures exist also to protect VoIP communications, notably SIP over TLS~\cite{SIPS} and SRTP~\cite{SRTP} which rely on cryptography. However, our considerations in this case differ from what we noted on printers.
Secure VoIP protocols not only require dedicated configurations on the phones but, notably, on the PBX server, e.g. Asterisk, which is necessary in support of the binding between caller and responder. At institutional level, such servers normally lie in the authorisation profiles of sysadmins only, who are also in charge of manually entering the long-term keys on the phones. An implication is that end-users cannot secure the entire configuration by themselves,  independently from the sysadmins.

We used the same social-network approach to somewhat confirm our hypothesis that secure VoIP protocols are still less common than traditional ones today, raising the significance of the Printjack attacks. 
However, we feel that the main preoccupation in this scenario is that no straightforward measures exist for a pair of end-users to protect their VoIP communications from a malicious sysadmin who may want to execute the Printjacks. Of course, the end-users could install their own Asterisk server from scratch then reset and reconfigure the phones entirely.
The sequel of this Section describes an alternative, novel measure to build a cryptographic tunnel over SIP.
Although it requires some skills, we tested that it transparently integrates with the institutional VoIP configuration.

\subsection{Novel measures}
As noted above, measures already exist to enable any sufficiently skilled end-user to secure their client to any (IPSec compliant) printer --- from anyone, including malicious sysadmins. Therefore, we recognise that printers do not require any further security measures.
The opposite applies to VoIP communications, hence this Section describes the technological setup that any pair of sufficiently skilled end-users could adopt to seamlessly secure their communications while still relying on the institutional PBX server that someone else administrates. The gist is to shelter each phone with a simple device capable to encipher the audio stream, such as a Raspberry PI. Precisely, we want to thwart Phonejack 2 and Phonejack 3, so the new measure will have to thwart the malicious crashing of the phones as well as call sniffing.

\begin{figure}[ht]
	\begin{center}
		\includegraphics[width=0.95\linewidth]{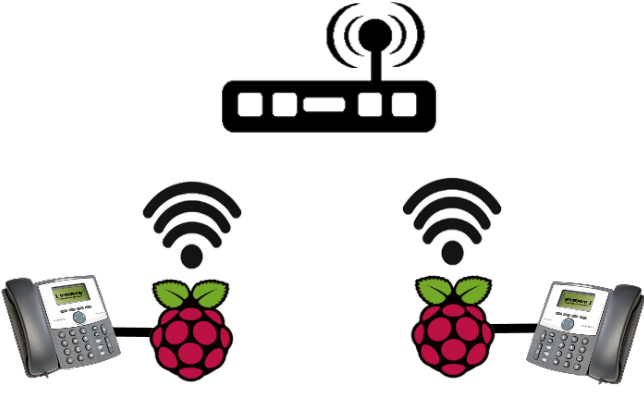}
	\end{center}
	\caption{An inexpensive network upgrade to support our protection measures}\label{fig:netArch}
\end{figure}

A recent model of Raspberry Pi is an inexpensive device that can be programmed to implement an encrypted tunnel between two VoIP phones. Precisely, these can be connected trough two Pis and a Wi-Fi bridge \cite{piwifibridge} as shown in Figure \ref{fig:netArch}. Precisely, each phone is connected to a Raspberry Pi through a wired Ethernet connection. Each Raspberry Pi will (have to) communicate via Wi-Fi with the other Raspberry Pi (because a Pi only has one Ethernet card).

For simplicity, static IP addresses are set via \textit{dhcpcd} and \textit{dnsmasq} tools. The Raspberry Pi connects to the Wi-Fi router, then the \textit{``/etc/dhcpcd.conf"} file must be modified by setting the interface, static IP address and the subnet. Then, the \textit{``/etc/dnsmasq.conf"} file must be modified to tell \textit{dnsmasq} how it should handle traffic. After that, it is time to activate the forwarding mode of the network card and configure Iptables as shown in Code \ref{code:iptablesrulenet} to effectively bridge Ethernet and Wi-Fi. The final step is to update the routing tables of each Pi.

\begin{lstlisting}[language=python, caption={Bridging Ethernet and Wi-Fi through Iptables}, numberstyle=\normalsize,  stepnumber=1, label={code:iptablesrulenet}]
iptables -t nat -A POSTROUTING -o wlan0 -j MASQUERADE

iptables -A FORWARD -i wlan0 -o eth0 -m state
--state RELATED,ESTABLISHED -j ACCEPT

iptables -A FORWARD -i eth0 -o wlan0 -j ACCEPT
\end{lstlisting}

Having upgraded the network, attention can be oriented to the actual attack protection measures. Since the VoIP flow must be modified, Iptables can redirect traffic and build 3 queues with which the three scripts will be associated (Code \ref{code:queues}). Specifically, queue 1 will be assigned with a dedicated anti-DoS script, then queues 2 and 3 respectively with scripts to encrypt and decrypt audio traffic.

\begin{lstlisting}[language=python, caption={Enqueuing SIP and RTP traffic through Iptables}, numberstyle=\normalsize,  stepnumber=1, label={code:queues}]
iptables -A FORWARD -p UDP -d PhoneAddress
--dport 5060 -j NFQUEUE --queue-num 1

iptables -A FORWARD -p UDP -s IPPhoneAddress
--sport rangeRTPport -j NFQUEUE --queue-num 2  

iptables -A FORWARD -p UDP -d PhoneAddress
--dport rangeRTPport -j NFQUEUE --queue-num 3   
\end{lstlisting}

\subsubsection{Countering Phonejack 2}

Code \ref{code:scriptantidos} shows our script to counter Phonejack 2. It analyses each packet received via the \texttt{get\_payload} function. It checks the file called \texttt{blacklist.txt}. If the analysed packet has been previously received, then it is discarded, otherwise it is accepted and marked as received in the file. The penultimate statement builds a queue via the \texttt{NetfilterQueue} library, while the last instruction connects the ID and the anti-DoS script to queue 1.

\begin{lstlisting}[language=python, caption={A Phonejack 2 protection measure in Python}, numberstyle=\normalsize,  stepnumber=1, label={code:scriptantidos}]
def antiDos(packet):
 pkt = IP(packet.get_payload())
 Flag=0
 with open('blacklist.txt') as f:
  if str(packet.get_payload()) in f.read():
  Flag=1
  if Flag == 1:
   packet.drop()
  else:
   packet.accept()
   f= open("blacklist.txt","a+")
   f.write(str(packet.get_payload()))
   f.close()
   Flag=0
nfqueue = NetfilterQueue()
nfqueue.bind(1, antiDos)
\end{lstlisting}

At this point, as shown in our video clip \cite{videophonejack}, each Raspberry Pi acts as a shield for a phone by filtering old packets from new ones while preserving voice communication.

\subsubsection{Countering Phonejack 3}

This Section implements our solution to encrypt and decrypt the audio stream without any significant overhead or additional latency to the call. Code \ref{code:scriptenc} shows the encryption script. It runs a few preliminary cryptographic operations and then executes the encryption function on the packet payload. Subsequently, it sends the encrypted packet via a socket and dequeues the packet. As with the anti-DoS script, also this script also be associated with a queue, in this case with the queue 2.

\begin{lstlisting}[language=python, caption={A Phonejack 3 protection measure in Python}, numberstyle=\normalsize,  stepnumber=1, label={code:scriptenc}]
def encrypt(packet):
 cipher_suite = Fernet(key)
 enc_vc=cipher_suite.encrypt(packet.get_payload())
 pkt = IP(packet.get_payload())
 MESSAGE = enc_vc
 sk = socket.socket(socket.AF_INET,socket.SOCK_DGRAM)
 sk.sendto(MESSAGE, (pkt[IP].dst, pkt[UDP].dport))
 packet.drop()
 
nfqueue = NetfilterQueue()
nfqueue.bind(2, encrypt)


def decrypt(packet):
 cipher_suite = Fernet(key)
 dec_vc=cipher_suite.decrypt(packet.get_payload())
 pkt = IP(packet.get_payload())
 MESSAGE = dec_vc
 sk=socket.socket(socket.AF_INET,socket.SOCK_DGRAM) 
 sk.sendto(MESSAGE,(pkt[IP].dst, pkt[UDP].dport))
 packet.drop()
 
nfqueue = NetfilterQueue()
nfqueue.bind(3, decrypt)
\end{lstlisting}

\begin{figure}[ht]
	\begin{center}
		\includegraphics[width=0.95\linewidth]{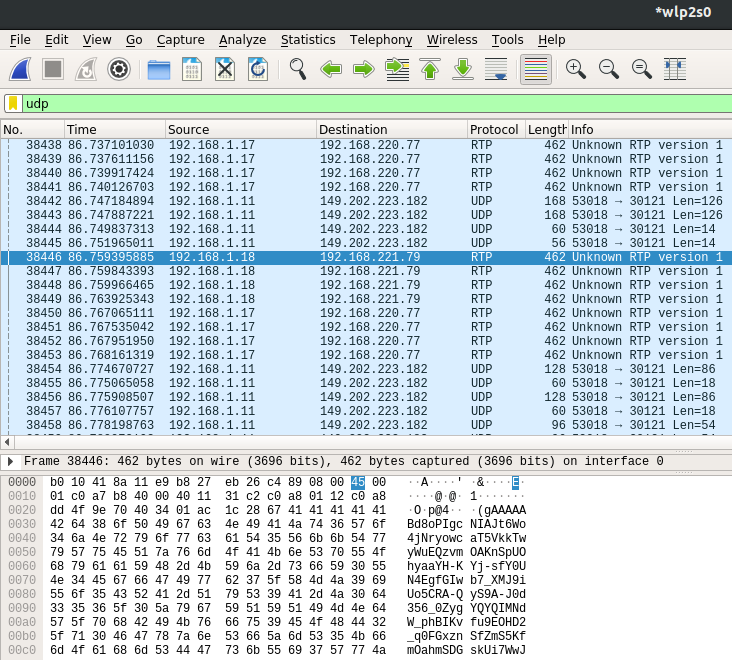}
	\end{center}
	\caption{A call sniffing attempt under our Phonejack 3 protection measure}\label{fig:sniffenc}
\end{figure}

It is now time to launch a sniffing job through Wireshark. This simulates a MITM who actively attempts call sniffing between the two phones. What the attacker would intercept is nothing but encrypted RTP traffic, as shown in Figure \ref{fig:sniffenc}.

The decryption script is similar to the encryption script but with two differences. One is the method invoked on the payload, which, in this case is \textit{decrypt}. The other one is the Iptables queue that is accessed, in this case queue number 3.

\begin{figure}[ht]
	\begin{center}
		\includegraphics[scale=0.35]{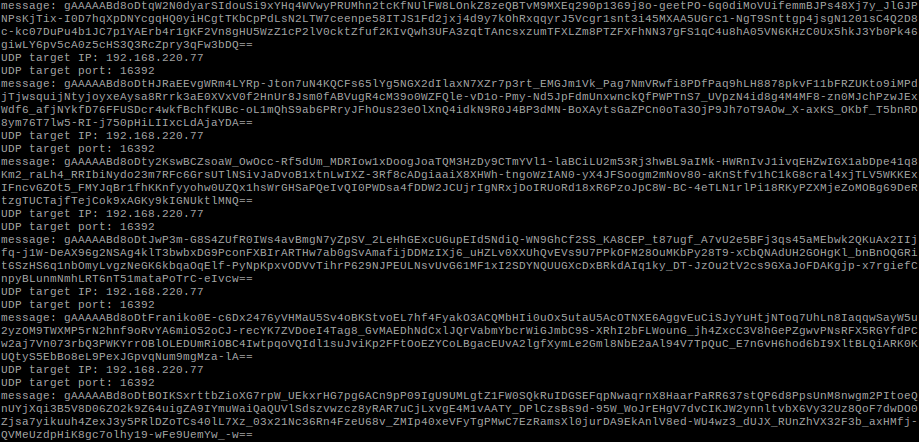}\\
		\vspace{0.25cm}
		\includegraphics[scale=0.35]{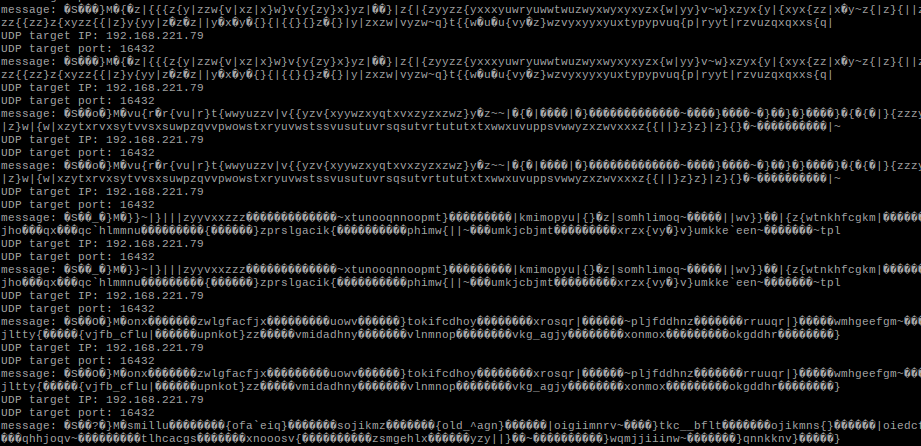}
	\end{center}
	\caption{Execution of our Phonejack 3 protection measure on a Raspberry Pi}\label{fig:trafficoconPI}
\end{figure}

Figure \ref{fig:trafficoconPI} shows how a Raspberry Pi manages traffic under our Phonejack 3 protection measure. The upper part of the Figure shows a phone running the encryption routine, and hence displays an encrypted RTP stream; conversely, the lower part shows a phone running the decryption routine, and hence displays unencrypted traffic.

These solutions can therefore be used in addition to the existing configurations, in such a way as to build on an institutional encrypted tunnel (which for example uses SIPS and SRTP), an additional tunnel that allows users to obtain greater confidentiality even towards a malicious system administrator.

\section{Conclusions}\label{sec:conclusions}
This article looks at network printers and VoIP phones focusing on real-world attacks using freeware and on corresponding protection measures that each individual user can put into place against anyone else. 
The Printjack family works against printers and the Phonejack family troubles phones. While attacks number 1 in each family are only conjectured, attacks number 2 realise innovative forms of denial of service and attacks number 3 demonstrate innovative or simplified data breaches. 
Extreme consequences might be Printjack 2 to fully exhaust an institution's ability to print documents out and Phonejack 2, if sprayed over all institutional phones, to cause building evacuation due to unbearable noise.
All data breaches are concerning, notably at European level, which is tightly regulated by the GDPR. In particular, Printjack 3 may subvert people's enrooted trust that printing out confidential documents is secure, and Phonejack 3 confirms the simplicity of call eavesdropping for insiders.
It must be remarked that, coherently with our research questions, these findings apply to real-world scenarios but only when the services are configured without security protocols.
Secure configurations are possible but we found out through social media queries as well as through Shodan searches that they are often neglected.

All attacks were carried out over a real-world network hosting the target services and, specifically, from inside the network, thus in a scenario in which the attacker is an insider. We did not try the attacks from the outside because we lack the 
necessary authorisations and indemnities to probe real-world networks from the outside. This is fully aligned with, and can be interpreted in our stated threat model. However, it implies that we can answer our research question RQ1 on the feasibility of attacking the two target services --- with a demonstrated `yes' --- only from the standpoint of insiders. The standpoint of outsiders can only be assessed in terms of a non-negligible likelihood, due to the number of worldwide ports that are visible from anywhere, that our attacks succeed.

By contrast, our demonstrated answer to our research question RQ2 on the feasibility of protecting the two services is fully affirmative. Each individual user may configure, subject to the necessary skills, IPSec-secured network printing and then empty their trust base from anyone else. The same can be achieved for VoIP calls by adopting our novel measure to build a cryptographic tunnel over SIP by means of two Raspberry Pis, a clear call for a \textit{by design and by default} integration of that technology with modern phones.

Our findings are significant because they impact two of the most common and vital institutional services. While the findings reconfirm the necessity of secure configurations, they also denounce that these may be rarely used. On one hand, innovative forms of denial of service and of data breach are found and demonstrated, on the other hand, innovative protective measures are designed and prototyped to secure VoIP calls while minimising the trust base.

Even the layman is increasingly accustomed at present to the necessity of securing their laptop because it might be hindered by attackers and the personal data it contains might be equally abused.
Our research supports the claim that similar security measures must be applied \textit{also} to network printing and to VoIP phones alike \textit{because} these might be hindered and the data they handle might be abused much the same way it can be done with laptops.
The same claim is likely to be applicable to other pervasive IoT devices, hence our current research is to investigate the extent to which our attacks would scale up to Wi-Fi cameras and what protection measures would be available or worth developing to fully secure them from virtually anyone beyond their legitimate users.

\bibliographystyle{abbrv}
\bibliography{iotbiblio}

\appendix

\section{A Primer on VoIP Protocols}\label{app:sip}
The essential protocols underlying VoIP are SIP and RTP.

\subsection{Session Initiation Protocol (SIP)}
Developed by the Internet Engineering Task Force (IETF), SIP consists in a telephone signaling protocol used to establish, modify and conclude VoIP phone calls \cite{SIP}. More precisely, it uses the UDP transport protocol with default port 5060 and has the following functions: i) authenticate, locate and acquire the audio coding preferences of the clients; ii) invite clients to participate in a session; iii) establish session connections; iv) carry a description of the session; v) manage any changes to the session parameters; vi) conclude telephone sessions.

The SIP protocol is based on a Client-Server system, in fact, generally there is a dedicated machine that plays this role, called SIP server. A client after being authenticated by the SIP server, can establish a connection with another client by the following steps:
\begin{enumerate}
	\item  The client sends this invitation message to the server. With this message, the client asks the server to establish a connection with the client indicated in the \texttt{user} parameter.\\
	\texttt{SIP INVITE:\textless sip : user@serverhost\textgreater}
	
	\item The server responds with a ``Trying" message to the client.
	
	\item The server forwards the invitation of the first phase to the recipient.
	
	\item The receiver sends a ``Trying" message to the server.
	
	\item The recipient sends a ``Ringing" message to the server, then, the server sends the same message to the sender. This message generates the typical call tone.
	
	\item When the recipient accepts the call, it sends an ``OK" message to the server. Subsequently, the server will notify the sender of the acceptance of the call.
	
	\item After the reception of the ACK message relating to the sixth phase, communication and transmission of the audio signal through the RTP protocol is established.
	
	\item To conclude the call, each party sends a ``BYE" message to the server, which will inform the other party.
\end{enumerate}

\subsection{Real-time Transport Protocol (RTP)}\label{app:rtp}
Also developed by the IETF, the Real-time Transport Protocol (RTP) complements SIP by providing end-to-end network transport functions suitable for real-time applications such as VoIP \cite{RTP}. RTP also uses UDP at the transport level. The RTP protocol does not have a default port, so different VoIP applications may choose a different port number. For example, Cisco SPA devices select the port number randomly from an established range of ports at each call. Once the RTP session is established, the clients use the audio coding specifications previously established thanks to the SIP protocol.

The information provided by the RTP protocol includes timestamps for synchronisation, sequence numbers useful in case of packet loss and the payload format that indicates the coded format of the data. Thanks to these fields in the RTP packets, the call can be reconstructed via Wireshark.
Unfortunately, the RTP protocol has some limitations, RTP does not control the quality of service (QoS), does not guarantee the delivery of packets and does not provide automatic retransmission of packets in case of loss.

\end{document}